\documentclass[12pt]{article} %\input epsf.tex
\usepackage{graphicx}
\newcommand{\be}{\begin{equation}}
\newcommand{\ee}{\end{equation}}
\newcommand{\bear}{\begin{eqnarray}}
\newcommand{\eear}{\end{eqnarray}}
\newcommand{\ba}{\begin{array}}
\newcommand{\ea}{\end{array}}

 \newcommand{\OO}{{\cal O}}
\begin{document}

\begin{titlepage}
\vfill
\begin{flushright}
{\normalsize IC/2010/012}\\
{\normalsize arXiv:1004.3541[hep-th]}\\
\end{flushright}

\vfill
\begin{center}
{\Large\bf  Electrified plasma in AdS/CFT correspondence }

\vskip 0.3in

{Bindusar Sahoo\footnote{\tt bsahoo@ictp.it}, Ho-Ung
Yee\footnote{\tt hyee@ictp.it}}

\vskip 0.15in

 {\it ICTP, High Energy, Cosmology and Astroparticle Physics,} \\
{\it Strada Costiera 11, 34151, Trieste, Italy}
\\[0.3in]

{\normalsize  2010}

\end{center}

\vfill

\begin{abstract}
We construct new gravity backgrounds holographic dual to neutral plasma with U(1) global symmetry
in the presence of constant electric field, considering its full back-reactions to the metric.
As the electric field and the induced current cause a net energy in-flow to the system, the plasma is continually heated up and the corresponding gravity solution has an expanding horizon.
After proposing a consistent late-time expansion scheme, we present analytic solutions in the scheme up to next-leading order, and
our solutions are new time-dependent solutions of 5D asymptotic AdS Einstein-Maxwell(-Chern-Simons) theory.
To extract dual CFT stress tensor and U(1) current from the solutions, we perform a rigorous holographic renormalization of Einstein-Maxwell-Chern-Simons theory including full back-reactions, which can in itself be an interesting addition to literatures.
As by-products, we obtain interesting modifications of energy-momentum/current Ward identities
due to the U(1) symmetry and its triangle anomaly.

\end{abstract}

\vfill

\end{titlepage}
\setcounter{footnote}{0}

\baselineskip 18pt \pagebreak
\renewcommand{\thepage}{\arabic{page}}
%\tableofcontents
\pagebreak

\section{Introduction }

AdS/CFT correspondence provides an important opportunity for both gravity and strongly coupled CFT  \cite{Maldacena:1997re}.
Many features in gravity that have been found before the correspondence find their nice physical re-interpretations as field theory phenomena in the dual CFT. Study of a strongly coupled CFT, especially its
time-dependent dynamics, in terms of conventional field theory techniques, is hampered by many intrinsic difficulties. AdS/CFT correspondence is one consistent framework that can provide
reliable information on some aspects of these strongly coupled gauge theories.

Finite temperature system, or plasma, seems particularly interesting to study in the framework \cite{Witten:1998zw,Policastro:2002se}, because of its possible relevance to quark-gluon plasma one observes in heavy-ion collisions which is believed to be strongly coupled \cite{Shuryak:2003ty}.
The expectation is that some finite temperature properties might depend weakly on details of microscopic theories, and be universal as long as interactions are strong \cite{Kovtun:2003wp,Buchel:2003tz}.
In the gravity side, a finite temperature plasma is conjecturally described by a black-hole, a space-time with horizon. Dynamics of the horizon seems to describe long wave-length hydrodynamics of the dual CFT plasma \cite{Policastro:2002se,Bhattacharyya:2008jc,Natsuume:2007ty,Iqbal:2008by}, while theory-specific non-hydrodynamic phenomena would involve the whole non-linear bulk space-time dynamics.
There has been much recent interest in these subjects; see Refs. \cite{Son:2007vk,Rangamani:2009xk,Eling:2010vr} for reviews.

One way of studying plasma is to perturb the system and observe its linear responses.
This gives quite a lot of useful information on the properties of the plasma, mostly related one way or the other to hydrodynamic properties.
AdS/CFT correspondence is equipped with formalisms which nicely fits to this particular need \cite{Herzog:2002pc}. They are boundary perturbation \cite{Gubser:1998bc} and holographic renormalization \cite{de Haro:2000xn}.
As the perturbations one imposes are small and treated linearly, one typically forgoes considering back-reactions to the metric.

A less studied aspect is going beyond linearized approximations and taking back-reactions to the gravity equations of motion. One typically needs this extension in order to study far-from-equilibrium dynamics of the plasma, such as early-time thermalization or black-hole formations \cite{Kovchegov:2007pq,Gubser:2008pc,Chesler:2008hg,Lin:2009pn,Bhattacharyya:2009uu,Chesler:2009cy,irina}. It is also conceptually more appealing to have a full complete solution of the system, rather than working in an approximation. Another reason for including back-reactions is to study effects to the energy-momentum tensor from other symmetries/operators of the theory. For example, external gauge potential coupled to a global symmetry of a CFT can give rise to  violation of conformal Ward identities. To rigorously compute such effects, one needs to consider back-reactions to the metric.

In this work, we initiate a study for one particular case that indeed needs full back-reactions
to the metric. Our situation is a CFT plasma with U(1) global symmetry in the presence of constant external electric field. Our model includes N=4 SYM with U(1) R-symmetry as a particular case.
Because CFT possesses excitations of arbitrary small energy scale, the applied electric field inevitably causes an induced current along its direction. Previous linearized study of this situation gives us a conductivity \cite{CaronHuot:2006te}(see Ref.\cite{Karch:2007pd} for one non-linear study), but our intention is to go beyond this simple linearized description.
Observe that the electric field and the induced current do work on the system and there is a net energy in-flow to the system,
\be
{d\epsilon \over dt} = \vec E\cdot \vec J\quad,
\ee
so that the plasma will be continually heated-up. To correctly describe what is happening in a more long-time basis, one necessarily has to take into account gravity back-reactions from the U(1) sector. Especially, a natural expectation is to have an expanding horizon corresponding to heating the plasma, which might be an interesting time-dependent gravity solution in itself.

The action we study is
\be
\left(16\pi G_5\right) {\cal L} = R+12-{1\over 4} F_{MN} F^{MN} - {\kappa\over 4\sqrt{-g_5}}\epsilon^{MNPQR} A_M F_{NP} F_{QR}\quad,
\ee
where $\kappa={1\over 3\sqrt{3}}$ for N=4 SYM.
Although it is not difficult to write down simplified equations of motion of the situation exploring relevant symmetries of the system, such as translation-invariance in the present work, it seems in general to be a hard task to find exact analytic solutions. Instead, we follow a less ambitious path by invoking a consistent late-time expansion scheme, so that one can solve easier sets of differential equations order by order. The idea is similar to the Janik-Peschanki's late-time expansion of boost-invariant plasma \cite{Janik:2005zt}, although black-hole horizon in our case is expanding rather than receding as in their case. We stress that identifying correct scaling variable as well as the right expansion parameter
that we do here for our case is not a trivial result.
Within the proposed scheme, we succeed in finding smooth analytic solutions up to next-leading order in the expansion. Although the results will be more and more complicated as one goes beyond higher orders,
there is no further conceptual block in our scheme.
Our solutions seem new to our knowledge.

Because our solutions are taking full back-reactions to the metric, one also needs a full holographic renormalization \cite{de Haro:2000xn} of the Einstein-Maxwell(-Chern-Simons) system including back-reactions. To our knowledge, there has not been complete analysis of this, and we intend to fill this gap in this work.
Particularly interesting by-products of our analysis are a set of modifications of conformal Ward identity as well as energy-momentum/current conservation laws. From these results, we finally obtain the CFT stress tensor and the current of our new solutions up to next-leading order.

It is worth of mentioning that a similar problem with a constant magnetic field, instead of an electric field, was studied by D'Hoker-Kraus some time ago \cite{D'Hoker:2009mm}, with a quite different motivation. A major difference in their case is the static nature of the solution, which simplifies analysis in a fundamental way. In order to be more complete, we also tried to search for a static solution with constant electric field without current to observe a singularity at a finite radius, which makes the solution unacceptable. We think this is physically understandable because the current will be inevitably induced and the energy in-flow must happen so that the geometry is necessarily time-dependent \footnote{We thank Andy O'Bannon for raising this question to us.}.

We can broadly divide our paper into three parts. In the first part (section 2 and section 3), we propose a consistent late time expansion scheme and subsequently solve the equations of motion in the particular scheme till the first leading order. The next part (section 4) of the paper is devoted to using a rigorous holographic renormalization approach to derive the modified trace anomaly and modified Ward identity in the presence of an external electric field and taking into account the full backreaction. This can be an interesting result on its own independent of the other results of the paper. In the third part (section 5), we use the solution derived in section 3 to extract the energy momentum tensor and current density and observe that the obey the modified Ward identity derived in section 4. We don't think our analysis is complete in any respect, but our intention is to report our current
findings which might be interesting to a further study on the subject. We finish our paper by giving a few open, future problems to extend the current analysis.

\section{Constructing gravity solution : late time expansion}

The situation we are aiming at is a time-dependent finite temperature plasma with a $U(1)$ global symmetry
of which a constant external electric field is applied to the system.
For simplicity, we will be considering a neutral plasma, and we leave the case of finite charge density
as a future direction.
We are interested in obtaining an explicit 5-dimensional time-dependent gravity solution that
is holographic dual to the situation. Note that time-dependence is an unavoidable feature in the presence of the electric field, because a current $\vec J$ is expected to arise in response to the electric field $\vec E$, and
there will be a net energy flow into the system proportional to $\vec{J}\cdot\vec{E}$, by which the plasma will be heated up. Consequently, one expects the dual gravity solution to have an increasing event horizon as time goes on.

As one expects the plasma at hand to be heated by the applied electric field, it is natural
to have a thermodynamic concept of temperature, at least in sufficiently late time regime.
The problem of finding an exact gravity solution with a constant electric field is an interesting subject itself, although it seems difficult to our eyes to handle analytically.
Instead, we intend to study a late time asymptotic solution invoking a systematic late time expansion that will be discussed shortly, and to solve the equations of motion order by order.
We have checked that our expansion scheme works consistently at a few lowest orders, and
it seems very likely that the scheme is consistent at all orders. The philosophy is similar to the case
of late time expansion of boost invariant plasmas \cite{Janik:2005zt},
while the difference is that boost invariant evolution is a cooling process and the event horizon is receding, contrary to our case of heating up.

To set the notation, let a homogeneous electric field point to $x^3$-direction with a constant magnitude $E$. This will fix the unique boundary condition of the bulk $U(1)$ gauge field at the UV boundary of 5-dimensional asymptotic AdS space we are constructing.
One advantage of having a neutral system is that while the electric field will induce a net current
$\vec{J}= J_3 \hat{x^3}$ along the direction of electric field, there won't be any net {\it momentum flow}
generated along the field. Intuitively this can be understood by the fact that a neutral plasma
has an equal amount of thermally excited, positively and negatively charged carriers that will move
in opposite ways under an external electric field, so that there will be no net momentum flow generated by the current carriers. More rigorously, a neutral plasma has a charge conjugation symmetry under which
the electric field and the current flips its sign while the momentum flow is left unchanged. Invariance under this charge conjugation requires the momentum flow to vanish.
The absence of momentum flow will greatly simplify our gravity solution ansatz: one can safely put the $(t,x^i)$ ($i=1,2,3$) components of the metric to zero. Using diffeomorphisms one can further choose the gauge where
\be
g_{rr}= g_{ri}=0 \quad,\quad i=1,2,3\quad,\quad g_{tr}=1\quad,
\ee
where $r$ is the 5'th holographic coordinate. Geometrically it is an affine coordinate of ingoing null geodesics in Eddington-Finkelstein coordinate.
After renaming $(x^1,x^2,x^3)=(x,y,z)$, the resulting gravity ansatz that is consistent with
a left-over $SO(2)$ symmetry of $(x^1,x^2)$ rotation is written as
\be
ds^2 = -r^2 a(r,t) dt^2 +2dr dt + r^2 b(r,t) dz^2
+r^2 c(r,t)\left(dx^2+dy^2\right)\quad,
\ee
with three unknown, yet to be determined functions $(a,b,c)$ which depend only on $r$ and the time $t$.
We adopt the Eddington-Finkelstein coordinate as it has been proven to be useful to
implement the correct causal boundary condition at the horizon by simple {\it regularity} of the metric components \cite{Bhattacharyya:2008jc}. It was also shown that late time expansion of boost invariant plasmas works consistently
at all orders in this coordinate \cite{Kinoshita:2008dq}, and we conjecture the same in our case too.
To have an asymptotic AdS space with flat 4-dimensional CFT metric at the boundary,
one needs to impose the UV boundary condition at $r\to\infty$
\be
(a,b,c) \to (1,1,1)\quad ,\quad r\to\infty\quad,\label{bdry1}
\ee
for all $t$.

Regarding an ansatz for the bulk gauge field, it is clear that one needs to turn on $A_z$ component only,
in the gauge $A_t=0$. The non-normalizable mode of $A_z$ encodes the external electric field by
the UV boundary condition
\be
A_z \to E t\quad,\quad r\to\infty\quad,
\ee
while its normalizable mode, which should be determined by solving equations of motion,
contains the induced current $J_z$ along the field direction.
Note that one will have to turn on $A_t$ as well in the case of charged plasma, which
would complicate the analysis.
Therefore, the gauge field of our interest can be put simply as
\be
A= A(r,t) dz \quad,
\ee
with one more function $A(r,t)$ whose UV boundary condition is
\be
A(r,t)\to E t \quad,\quad r\to\infty\quad.\label{bdry2}
\ee
With the above Ansatzs for the metric and the $U(1)$ gauge field, the equations of motion
become a set of partial differential equations of four unknown functions $(a,b,c,A)$ with the specified boundary conditions. With a lack of further intuition to these differential equations in finding an exact solution, we will instead invoke a systematic late time expansion scheme that allows us to solve them
order by order analytically.

To identify the correct scaling variable and expansion parameter, we first observe that
the applied electric field becomes more {\it perturbative} to the plasma as it gets hotter as time goes on,
because due to conformal symmetry what matters is an effective dimensionless strength of
the electric field, $E\over T^2$, where $T$ is the temperature. As $T$ increases one therefore expects
that adiabatic linear response theory will describe the system more accurately.
Sub-leading effects coming from non-linear responses will be suppressed at late time regime
by powers of an expansion parameter $\left( E\over T^2\right)^2$ where the square is due to the charge conjugation symmetry.
In linear response theory of an electric field applied to an adiabatic plasma of temperature $T$,
the induced current is proportional to the electric field and is given by
\be
J_z= \sigma(T) E\quad,
\ee
where $\sigma(T)$ is the electrical conductivity at temperature $T$ of a neutral plasma.
 Conformal invariance dictates that $\sigma(T)\propto T$.
Then the energy in-flow due to the applied electric field will be
\be
{d\epsilon \over dt}\sim T^3 {dT\over dt}=\vec{J}\cdot\vec{E} \sim E^2 \cdot T\quad,
\ee
where we have used the fact that a conformal plasma energy density $\epsilon$ has $\sim T^4$ behavior.
The easy solution of the above differential equation for $T$ is
\be
T\sim E^{2\over 3}\cdot t^{1\over 3}\quad,
\ee
for the leading late time behavior of the plasma at large $t$.
In the gravity side, temperature $T$ roughly corresponds to the location of black-hole horizon, that is,
$r_H\sim T$, and in the leading dual gravity solution one may expect to have a horizon moving
towards the UV region as $r_H\sim t^{1\over 3}$ in the late time regime.
This naturally motivates a scaling variable
\be
u\equiv {r \over t^{1\over 3}}  \quad,
\ee
in terms of which the leading solution will look "static", and there will be sub-leading
corrections to this leading solution by suitable negative powers of $t$.
A natural candidate for this sub-leading expansion parameter is the previously mentioned non-linearity
\be
\left( E\over T^2\right)^2 \sim t^{-{4\over 3}}\quad,\label{small}
\ee
and we conjecture that this late time expansion scheme works consistently order by order.
Notice also that the temperature at late time grows more and more slowly
\be
{dT\over dt} \sim t^{-{2\over 3}} \to 0 \quad,
\ee
so that our adiabatic analysis in the above seems self-consistent.

More explicitly, our first naive proposal, which in fact needs a refinement that will be discussed shortly, would be
\be
(a,b,c)(r,t) = \sum_{n\ge 0}^\infty \left(a_n(u),b_n(u),c_n(u)\right)t^{-{4n\over 3}}\quad,\quad
u\equiv {r \over t^{1\over 3}} \quad,\label{exp1}
\ee
for the functions entering our metric ansatz with the boundary conditions as $u\to\infty$,
\be
(a_0(u),b_0(u),c_0(u))\to (1,1,1)\quad,\quad (a_{n}(u),b_n(u),c_n(u))\to(0,0,0) \quad ,\quad n\ge 1\quad.
\ee
One needs a slightly more care for $A(r,t)$ in the gauge field. The leading expectation for the current $J_z$ is
\be
J_z\sim \sigma(T) E \sim T\cdot E \sim t^{1\over 3}\quad,
\ee
and this current sits in the normalizable ($\sim {J_z\over r^2}$) mode of $A_z$, so that the leading
late time behavior of $A(r,t)$ near UV boundary should be
\be
A(r,t)\sim E t +{t^{1\over 3}\over r^2} +\cdots = E t+{1\over t^{1\over 3} u^2}+\cdots\quad.\label{gaugeexp}
\ee
This suggests the following late time expansion for the function $A(r,t)$,
\be
A(r,t)= Et + {1\over t^{1\over 3}}\sum_{n\ge 0}^\infty j_n(u) t^{-{4n\over 3}}\quad,\label{exp2}
\ee
with the boundary condition,
\be
j_n(u)\to 0\quad,\quad n\ge 0\quad,\quad u\to\infty\quad.
\ee

The validity of the expansion scheme (\ref{exp1}) and (\ref{exp2}) can be tested by performing a few lowest order terms in the equations of motion to see whether one finds a consistent solution order by order.
Although one finds a good solution at leading zero'th order which is presented explicitly in the next section, there seems to appear an inconsistency at subsequent sub-leading orders, which necessitates an improvement of our first guess above. One however finds that the equations of motion indeed have
the right expansion parameter (\ref{small}), so that our arguments for $(u,t^{-{4\over 3}})$ seem more or less correct up to a possible subtlety yet to be identified.

One observes a hint for the resolution in a near boundary analysis of general solutions of our equations of motion, which one typically performs in the procedure of holographic renormalization \cite{de Haro:2000xn}. We will discuss more details about holographic renormalization in section~\ref{sectionhol}.
Expanding $(a,b,c,A)(r,t)$ near $r\to\infty$ with the boundary conditions (\ref{bdry1}) and (\ref{bdry2}),
one finds that the equations of motion dictate
\bear
a(r,t)&=& 1+{a^{(1)}(t)\over r}+\left({1\over 4}\left(a^{(1)}(t)\right)^2 -\partial_t a^{(1)}(t)\right) {1\over r^2}
+{a^{(4)}(t)\over r^4}-{E^2\over 6}{\log r\over r^4}+\cdots\,,\nonumber\\
b(r,t)&=& 1+{a^{(1)}(t)\over r}+{\left(a^{(1)}(t)\right)^2\over 4} {1\over r^2}
+{b^{(4)}(t)\over r^4}-{E^2\over 6}{\log r\over r^4}+\cdots\,,\nonumber\\
c(r,t)&=& 1+{a^{(1)}(t)\over r}+{\left(a^{(1)}(t)\right)^2\over 4} {1\over r^2}
+\left(-{1\over 2}b^{(4)}(t)-{E^2\over 24}\right){1\over r^4}+{E^2\over 12}{\log r\over r^4}+\cdots\,,\nonumber\\
A(r,t)&=& E t + {E\over r}+ \left(j^{(2)}(t)-{1\over 2}E a^{(1)}(t)\right){1\over r^2} + 0\cdot {\log r\over r^2} +\cdots\,,\label{nbdry}
\eear
with four undetermined functions $(a^{(1)},a^{(4)},b^{(4)},j^{(2)})$ of time with one constraint
equation
\be
\partial_t a^{(4)}(t)={2\over 3} E \,j^{(2)}(t)\quad,\label{curcons}
\ee
which will be shown to lead to energy conservation law.
The function $a^{(1)}(t)$ is in fact spurious as it arises from a simple coordinate reparameterization
\be
r\to r+{1\over 2} a^{(1)}(t)\quad,
\ee
and subsequently one safely ignores it from now. $(a^{(4)},b^{(4)},j^{(2)})$ encode the
expectation values of energy-momentum and $U(1)$ current in the dual CFT whose precise formulae will be given in section \ref{sectionhol2} later, but what is interesting for us at the moment is the appearance
of several $\log r\over r^4$ terms in the metric functions $(a,b,c)$ due to a backreaction of the external electric field $E$. In the case of pure Einstein gravity, this $\log r\over r^4$ term is given by
certain curvature tensors of boundary CFT metric which is related to the conformal anomaly, and
in our case of flat CFT metric it would have vanished.  In the presence of external gauge potential
that couples to $U(1)$ current, it gets additional contributions from field strengths of the external gauge field as our results indicate.
We stress that the presence of $\log r\over r^4$ terms is {\it necessary} for a consistent solution of equations of motion. Moreover, as the equations of motion are non-linear, one generically expects higher powers like
\be
\left( \log r\over r^4\right)^m\quad,\quad m\ge 1\quad,
\ee
to appear in further sub-leading expansions of actual solutions.

What we observe in the above paragraphs, (which we will discuss in more details in our section on holographic renormalization), is that in the holographic renormalization procedure, presence of log terms in the near boundary solutions is inevitable in the presence of non trivial electric field and non trivial background CFT metric. The late time expansion scheme is supposed to approximate the full bulk solution (which includes the near boundary behaviour as well) at late times and hence it should commensurate well with this observation of near boundary behaviour of the fields. This calls for refining the late time ansatz by addition of suitable log terms. In the next few paragraphs we will identify such terms.

As discussed in the previous paragraphs, in terms of our scaling and expansion variables $(u\equiv r t^{-{1\over 3}},t^{-{4\over 3}})$,
a consistent expansion scheme of $(a,b,c,A)$ should be able to reproduce these logarithmic structures near
$r\to\infty$. In terms of $(u,t^{-{4\over 3}})$, one writes
\be
{\log r\over r^4}={\log u\over u^4} t^{-{4\over 3}}+{1\over 3 u^4}{\log t \cdot t^{-{4\over 3}}}\quad,
\ee
and one observes that the first piece can easily be captured by next leading terms of the previous naive expansion scheme, while the second piece with $\log t$ can never be taken care of because there is no way
to introduce $\log t$ in the naive expansion scheme. The easiest way to remedy the problem is to invoke
a refined expansion scheme including powers of $\left(\log t \cdot t^{-{4\over 3}}\right)$
in addition to usual powers of $t^{-{4\over 3}}$. We conjecture and confirm at a few lowest orders that
$\log t$ indeed comes in a combination $\left(\log t \cdot t^{-{4\over 3}}\right)$, so that multiple logs are suppressed by at least same number of powers of $t^{-{4\over 3}}$.
This naturally leads to our proposal of late time expansion scheme
\bear
(a,b,c)(r,t) = \sum_{(n,m)\ge(0,0)}\left(a_{(n,m)}(u),b_{(n,m)}(u),c_{(n,m)}(u)\right)
t^{-{4\over 3}(n+m)}\left(\log t\right)^m\,,\, u\equiv {r\over t^{1\over 3}}.\label{newexp1}
\eear
One arrives at a similar conclusion for $A(r,t)$, too. From the equations of motion, one finds that
there appears a logarithmic term in the near boundary analysis starting at $\log r \over r^3$ whose
coefficient is roughly of order of $\partial_t j^{(2)}(t)$. As we know that $j^{(2)}(t)\sim t^{1\over 3}$
(see eq.(\ref{gaugeexp})), we have a term of
\be
t^{-{2\over 3}}{\log r\over r^3}\sim {1\over t^{1\over 3}}\left({\log u\over u^3} t^{-{4\over 3}}
+{1\over 3 u^3} {\log t \cdot t^{-{4\over 3}}}\right)\quad,
\ee
which guides us to the following expansion
\be
A(r,t)= Et + {1\over t^{1\over 3}}\sum_{(n,m)\ge (0,0)} j_{(n,m)}(u) t^{-{4\over 3}(n+m)}\left(\log t\right)^m \quad.\label{newexp2}
\ee
At each order $N\ge 0$, all functions of total order $n+m=N$ will enter the equations of motion in general.
Note that while the number of unknown functions increases as $N$, there are also $N$ number of equations to solve
because the equations of $(\log t)^m$ factors with $m\le N$ are all independent.
Eqs.(\ref{newexp1}) and (\ref{newexp2}) are our proposal for a consistent late time expansion
of the gravity background corresponding to heating a plasma by an external electric field.

We finish this section by commenting one caveat.
The equations of motion have an obvious time-translation symmetry of $t\to (t+t_0)$, so that from any solution one gets another by simply replacing $t$ with $(t+t_0)$. In particular our scaling variables $(u, t^{-{4\over 3}})$ maps to
\be
\left(u, t^{-{4\over 3}}\right)\to \left(u\cdot\left(1+t_0 t^{-1}\right)^{-{1\over 3}}, t^{-{4\over 3}}\cdot\left(1+t_0 t^{-1}\right)^{-{4\over 3}}\right)\quad,
\ee
under this transformation. Starting from our expansion (\ref{newexp1}) and (\ref{newexp2}), this transformation would generate a new series of {\it sub-leading} terms with additional integer powers of $t^{-1}$, so that a more general series that also covers this freedom seems to be
\be
\sum_{(n,m,p)\ge(0,0,0)} t^{-{4\over 3}(n+m)}t^{-p}\left(\log t\right)^m\quad.
\ee
However, it is natural to conjecture that these additional terms with $t^{-p}$ are rigidly determined with one free integration constant $t_0$, and
reorganize themselves to our expansion of (\ref{newexp1}) and (\ref{newexp2}) with $t$ being $(t+t_0)$ instead, so that one can always return to our proposal (\ref{newexp1}) and (\ref{newexp2}) by suitable
translation of time coordinate $t$. We leave a proof of this claim to the future.

\section{Leading and next leading order gravity solution}

In this section,
we provide a consistent analytic solution in our late time expansion scheme introduced in the previous section up to the order $n+m\le 1$. We expand the Einstein equation ${\cal E}^M_{\,\,\,\,N}=0$ and the Maxwell equation ${\cal M}_N=0$ in powers of $(t^{-1},\log t)$, and let us denote the coefficient of $t^{-\alpha}\left(\log t\right)^\beta$ by a superscript $(\alpha,\beta)$. Then
${\cal E}^{t(0,0)}_{\,\,\,\,t}$, ${\cal E}^{t({1\over 3},0)}_{\,\,\,\,u}$, ${\cal E}^{u({1},0)}_{\,\,\,\,t}$, ${\cal E}^{u({0},0)}_{\,\,\,\,u}$, ${\cal E}^{z({0},0)}_{\,\,\,\,z}$, ${\cal E}^{x({0},0)}_{\,\,\,\,x}$ , and ${\cal M}^{({1\over 3},0)}_z$ provide complete
differential equations for the leading order functions of $(n,m)=(0,0)$, and one also checks that
these equations are not affected by sub-leading functions of $(n,m)>(0,0)$.
As these leading order equations are {\it non-linear}, there is no systematic way to solve them, and one generally needs an educated guess. Once the leading order functions are found however, the differential equations for sub-leading functions in further expansions of equations of motion are {\it linear} differential equations, and one can solve them order by order systematically.

Because one expects the leading solution to be a black-hole solution whose event horizon would look static in the scaling coordinate $u$, one can try the following ansatz guided by the static black-hole solution,
\be
a_{(0,0)}(u)=1-{w^4\over u^4}\quad,\quad b_{(0,0)}(u)=c_{(0,0)}(u)=1\quad,
\ee
with some constant $w$ to be determined later, and one indeed finds that this solves almost all the above differential equations, namely,  ${\cal E}^{t(0,0)}_{\,\,\,\,t}$, ${\cal E}^{t({1\over 3},0)}_{\,\,\,\,u}$, ${\cal E}^{u({0},0)}_{\,\,\,\,u}$, ${\cal E}^{z({0},0)}_{\,\,\,\,z}$, and ${\cal E}^{x({0},0)}_{\,\,\,\,x}$ are satisfied by this ansatz. The remaining two equations ${\cal E}^{u({1},0)}_{\,\,\,\,t}$ and ${\cal M}^{({1\over 3},0)}_z$ become then a first/second order differential equation for the remaining $j_{(0,0)}(u)$ respectively. Explicitly, ${\cal E}^{u({1},0)}_{\,\,\,\,t}$ gives us
\be
j_{(0,0)}'(u)=-{u\left(E^2 u-4 w^4\right)\over E\left(u^4-w^4\right)}\quad,\label{j001st}
\ee
while ${\cal M}^{({1\over 3},0)}_z$ provides a second order differential equation
\be
u\left(u^4-w^4\right)j_{(0,0)}''(u)+\left(3u^4+w^4\right)j_{(0,0)}'(u)+E u^2 =0\quad.\label{j002nd}
\ee
It is simple to integrate (\ref{j001st}), but notice that the integrand on the right-hand side has a simple pole at the event horizon $u=w$ which would cause a logarithmic singularity of $j_{(0,0)}$, unless
this pole is removed by a suitable choice of constant $w$ such that the numerator also vanishes at $u=w$.
This uniquely fixes $w$ to be
\be
w=\left(E\over 2\right)^{2\over 3}\quad,
\ee
and the integration of (\ref{j001st}) with the UV boundary condition $j_{(0,0)}(\infty)=0$ results in
\be
j_{(0,0)}(u)=\left(E\over 2\right)^{1\over 3}\left({\pi\over 2} - {\rm tan}^{-1}\left(\left(2\over E\right)^{2\over 3} u\right)+{1\over 2}\log\left(\left(\left(E\over 2\right)^{2\over 3}+u\right)^2\over\left(\left(E\over 2\right)^{4\over 3}+u^2\right)\right)\right)\quad.
\ee
It is then nicely checked that this solution also solves the remaining equation (\ref{j002nd}), which is a non-trivial test for consistency of the solution.
In summary, {\it horizon regularity} and UV boundary condition uniquely determine the leading order solution without any
further integration constant. This feature will in fact be true in subsequent sub-leading higher-order solutions too.

The differential equations for next leading order functions of $(n,m)=(1,0)$ or $(0,1)$
are provided by the same set of Einstein equations and Maxwell equation as above with $\alpha\to\alpha+{4\over 3}$ and $\beta=0$ or $1$.
As they are linear differential equations, solving them has no conceptual problem while it is
algebraically quite cumbersome to present, so we only sketch the procedure and simply present the results.
One first observes that $\left({\cal E}^{t({5\over 3},0)}_{\,\,\,\,u},{\cal E}^{t({5\over 3},1)}_{\,\,\,\,u}\right)$ are second order differential equations for the combinations $\left(b_{(1,0)}+2c_{(1,0)},b_{(0,1)}+2c_{(0,1)}\right)$ respectively, whose integration gives us
\bear
b_{(1,0)}+2c_{(1,0)}&=&\left(2\over E\right)^{2\over 3}\left({\pi\over 2} - {\rm tan}^{-1}\left(\left(2\over E\right)^{2\over 3} u\right)+{1\over 2}\log\left(\left(\left(E\over 2\right)^{2\over 3}+u\right)^2\over\left(\left(E\over 2\right)^{4\over 3}+u^2\right)\right)\right)\nonumber\\
&+&{{\pi\over 2}-2-\tan^{-1}\left(\left(2\over E\right)^{2\over 3} u\right)\over u}+{C_1\over u}\quad,\nonumber\\
b_{(0,1)}+2c_{(0,1)}&=&{\tilde C_1\over u}\quad,\label{b+2c}
\eear
where we already fixed one integration constant by UV boundary condition while the remaining integration constants $(C_1,\tilde C_1)$ will be discussed shortly.
Next one finds that the equations $({\cal E}^{z({4\over 3},0)}_{\,\,\,\,z}+2{\cal E}^{x({4\over 3},0)}_{\,\,\,\,x},{\cal E}^{z({4\over 3},1)}_{\,\,\,\,z}+2{\cal E}^{x({4\over 3},1)}_{\,\,\,\,x})$
provide first order differential equations of $(a_{(1,0)},a_{(0,1)})$ respectively after using the above (\ref{b+2c}). It is not difficult to integrate them to have
\bear
a_{(1,0)}&=&{C_1\over 3u}\left(1+\left(E\over 2\right)^{8\over 3}{1\over u^4}\right)+{C_2\over u^4}+{1\over 3u^4}\Bigg[-\left(E\over 2\right)^{2\over 3}u^2-{2\over u}\left(E\over 2\right)^{8\over 3}\nonumber\\&+&{1\over u}\left(\left(E\over 2\right)^{8\over 3}+u^4\right)\left({\pi\over 2}-\tan^{-1}\left(\left(2\over E\right)^{2\over 3} u\right)\right)
-\left(E\over 2\right)^2\log\left(E^{4\over 3}+2^{4\over 3}u^2\right)\Bigg]\quad,\nonumber\\
a_{(0,1)}&=&{\tilde C_1\over 3u}\left(1+\left(E\over 2\right)^{8\over 3}{1\over u^4}\right)+{\tilde C_2\over u^4}\quad,\label{a}
\eear
with two more integration constants $(C_2,\tilde C_2)$.
Then, with the above results of (\ref{b+2c}) and (\ref{a}), one can check that the equations  $({\cal E}^{t({4\over 3},0)}_{\,\,\,\,t},{\cal E}^{t({4\over 3},1)}_{\,\,\,\,t})$ and $({\cal E}^{u({4\over 3},0)}_{\,\,\,\,u},{\cal E}^{u({4\over 3},1)}_{\,\,\,\,u})$ are automatically satisfied.
Next using (\ref{b+2c}) and (\ref{a}), the equations $({\cal E}^{z({4\over 3},0)}_{\,\,\,\,z},{\cal E}^{z({4\over 3},1)}_{\,\,\,\,z})$
become second order differential equations for $(c_{(1,0)},c_{(0,1)})$ respectively, whose first integration produces
\bear
c_{(1,0)}'(u)&=&-{C_1\over 3u^2}+{1\over 3u^2}\Bigg[\left(2-{\pi\over 2}\right)+\tan^{-1}\left(\left(2\over E\right)^{2\over 3}u\right)\nonumber\\
&+&{u\left(2u^3-\left(E\over2\right)^{2\over 3}u^2 +{E^2\over 2}\log\left(E^{4\over 3}+2^{4\over 3}u^2\right)-C_3\right)\over\left(\left(E\over 2\right)^{8\over 3}-u^4\right)}\Bigg]\quad,\nonumber\\
c_{(0,1)}'(u)&=&-{\tilde C_1\over 3u^2}-{\tilde C_3\over 3u}{1\over \left(\left(E\over 2\right)^{8\over 3}-u^4\right)}\quad,\label{cprime}
\eear
with integration constants $(C_3,\tilde C_3)$ to be determined shortly. Note that the right-hand sides of (\ref{cprime}) have
simple poles at the horizon $u=w=\left(E\over 2\right)^{2\over 3}$ unless $(C_3,\tilde C_3)$ are chosen to remove the singularity.
This regularity condition determines $(C_3,\tilde C_3)$ uniquely as
\be
C_3=\left(E\over 2\right)^2\left(1+2\log\left(2E^{4\over 3}\right)\right)\quad,\quad \tilde C_3=0\quad.
\ee
If one wishes, one can further integrate (\ref{cprime}) with UV boundary condition  to find $(c_{(1,0)},c_{(0,1)})$ explicitly.
For $c_{(0,1)}$, the result is simply
\be
c_{(0,1)}={\tilde C_1\over 3u}\quad,
\ee
while the result for $c_{(1,0)}$ looks quite complicated and we skip writing it down explicitly by presenting only
its near boundary asymptotic as
\be
c_{(1,0)}\sim {C_1\over 3u}+{E^2\over 36 u^4}\left(3\log u-2\log E +{1\over 2}\log 2 -{1\over 4}\right)+\cdots\quad,\quad u\to\infty\quad,
\ee
for a later convenience.
Combined with this, (\ref{b+2c}) and (\ref{a}) give us complete next leading-order solutions of $(a,b,c)$ of $(n,m)=(1,0)$ and $(0,1)$ with
integration constants $(C_{1,2},\tilde C_{1,2})$ yet to be determined.

The remaining final equations to solve, namely
$({\cal E}^{u({7\over 3},0)}_{\,\,\,\,t},{\cal E}^{u({7\over 3},1)}_{\,\,\,\,t})$ and $({\cal M}^{({5\over 3},0)}_z,{\cal M}^{({5\over 3},1)}_z)$ provide first/second order differential equations for
$(j_{(1,0)},j_{(0,1)})$ respectively. The first order equations $({\cal E}^{u({7\over 3},0)}_{\,\,\,\,t},{\cal E}^{u({7\over 3},1)}_{\,\,\,\,t})$ take the following structure
\be
\left(u^4-\left(E\over 2\right)^{8\over 3}\right)j_{(1,0)}'(u)=F_{(1,0)}(u)\quad,\quad
\left(u^4-\left(E\over 2\right)^{8\over 3}\right)j_{(0,1)}'(u)=F_{(0,1)}(u)\quad,\label{jprime}
\ee
where $(F_{(1,0)},F_{(0,1)})$ are some complicated combinations of $(a,b,c)$ and $j_{(0,0)}$ so that they depend on the integration constants $(C_{1,2},\tilde C_{1,2})$. To avoid logarithmic singularity
at the horizon $u=w=\left(E\over 2\right)^{2\over 3}$, one has to require $F_{(1,0),(0,1)}(w)=0$. By explicit computations, one finds that $F_{(1,0),(0,1)}(w)$ do not involve $(C_1,\tilde C_1)$ so that
this regularity uniquely determines $(C_2,\tilde C_2)$ only. Explicitly one has
\bear
F_{(1,0)}(w)={E^{5\over 3}\over 3\,2^{8\over 3}}\left(1-\log2 -4\left(\log2\right)^2 -{4\over 3}\log E +{12\over E^2}\left(C_2-3\tilde C_2\right)\right)\,,\, F_{(0,1)}(w)={\tilde C_2\over 2^{2\over 3} E^{1\over 3}}\,,\nonumber
\eear
so that this fixes $(C_2,\tilde C_2)$ uniquely as
\be
C_2={E^2\over 12}\left(\log2-1+4\left(\log2\right)^2 +{4\over 3}\log E\right)\quad,\quad\tilde C_2=0\quad.
\ee
After fixing $(C_2,\tilde C_2)$, it is not difficult to integrate (\ref{jprime}) to obtain $j_{(0,1)}$ as
\be
j_{(0,1)}(u)=-\tilde C_1{E\over 6}{u\over\left(u^3+\left(E\over 2\right)^{2\over 3}u^2+\left(E\over 2\right)^{4\over 3}u+\left(E\over 2\right)^2\right)}\quad,
\ee
while the expression for $j_{(1,0)}$ is too complicated to present here.
For our later purpose, we present explicitly only the near UV behavior of $j_{(1,0)}$,
\be
j_{(1,0)}(u)\sim {E\over 12 u^2}\left(1-2C_1\right)+{\cal O}\left(u^{-3}\right)\quad.
\ee
It is not difficult to check also that these results for $j_{(1,0)}$ and $j_{(0,1)}$ from (\ref{jprime}) satisfy the second order differential equations $({\cal M}^{({5\over 3},0)}_z,{\cal M}^{({5\over 3},1)}_z)$, which is a consistent check for the framework.

Finally, the remaining integration constants $(C_1,\tilde C_1)$ can easily be identified as spurious modes coming from a simple coordinate re-parameterization
\be
r\to r+{C_1\over 6t}+{\tilde C_1\log t\over 6 t}\quad,
\ee
or equivalently
\be
u\to u+{C_1\over 6t^{4\over 3}}+{\tilde C_1\log t\over 6 t^{4\over 3}}\quad,
\ee
which is a left-over gauge freedom in our coordinate system. See $a^{(1)}(t)$ in eq.(\ref{nbdry}).
Therefore one can safely set them zero, and one is left with the unique next leading solution presented above with all the integration constants completely determined by regularity at the horizon.

\section{Holographic renormalization of Einstein-Maxwell-Chern-Simons theory }\label{sectionhol}

The purpose of this section is to perform a rigorous holographic renormalization of Einstein-Maxwell-Chern-Simons theory,
including full back reaction to the metric. Our motivation is two-fold.
First, to our knowledge there has been no complete analysis on a holographic renormalization of Einstein-Maxwell theory taking into account back reactions to the metric, and it seems useful to clear
the situation including Chern-Simons term too.
Our second purpose is to extract interesting physical observables from our new gravity background in the previous section, such as energy-momentum and $U(1)$ current, using the rigorous results of holographic renormalization. The necessity of a careful holographic renormalization is due to the presence of external electric field which breaks the underlying conformal symmetry, and one naturally expects a violation of
conformal Ward identity $T^\mu_{\,\,\,\,\mu}=0$. This hints to a possible non-trivial effect
to the energy-momentum tensor from the electric field, and at least conservatively, one needs to follow rigorous steps of
holographic renormalization including back reactions to the metric to correctly identify it.

As we will formulate regularization and counter-terms in a covariant way, we can choose to
work in the standard (and simplest) Fefferman-Graham coordinate, although our gravity background
is written in an Eddington-Finkelstein like coordinate.
After identifying {\it covariant} expressions of counter-terms, one can safely apply the results to our Eddington-Finkelstein coordinate (or any other reasonable coordinate) to extract renormalized finite values of physical observables.
\subsection{Near boundary solution to the equations of motion}
{\it A la} Ref.\cite{de Haro:2000xn}, one starts with near UV boundary behaviors of the metric and $U(1)$ gauge field which are
dictated by the equations of motion,
\bear \label{holren1}
ds^2&=& G_{MN}dx^M dx^N={d{\rho^2}\over {4{\rho}^2}}+ {1 \over \rho}g_{\mu\nu}(x,\rho)dx^{\mu}dx^{\nu}\quad, \nonumber \\
A_\mu &=& A_\mu^{(0)}+A_\mu^{(2)}\rho+B_\mu^{(2)}\rho \log \rho +\cdots\quad,
\eear
with
\be
g_{\mu\nu}(x,\rho)= g^{(0)}_{\mu\nu}+ g^{(2)}_{\mu\nu}\rho
+g^{(4)}_{\mu\nu}\rho^{2}+h^{(4)}_{\mu\nu}\rho^{2}\log \rho +\cdots \quad,
\ee
where $g^{(0)}$ and $A^{(0)}$ are {\it Dirichlet} boundary conditions that are given by hand as external CFT data such as background 4-dimensional metric and an external potential coupling to $U(1)$ current.
As in the case of pure Einstein gravity and Maxwell theory without back reaction, the terms $g^{(2)}$, $h^{(4)}$ and $B^{(2)}$ are expected to be completely fixed by equations of motion in terms of these
external CFT data, $g^{(0)}$ and $A^{(0)}$. The contributions from $A^{(0)}$ especially to $g^{(2)}$ or $h^{(4)}$ would be a new aspect coming from back reactions, whose presence can already be seen in our previous near boundary behaviors (\ref{nbdry}).
The next terms of $g^{(4)}$ and $A^{(2)}$ are {\it dynamical}, and not completely determined by near boundary equations of motion,  but
are constrained by them. Finite, renormalized CFT observables are given in terms of these coefficients, and the constraint equations from equations of motion typically give us Ward-identities for these observables.
As mentioned before, we are interested in these Ward-identities too.

The action for the Einstein-Maxwell-CS theory is
\bear \label{holren2}
S &=&{1\over {16\pi G_{5}}}\left[\int_{M} d^4 x d\rho \sqrt{-G}\left(R[G]+12- {1\over 4} F_{MN}F^{MN}\right) - {\kappa \over 4} \epsilon^{MNPQR}A_{M}F_{NP}F_{QR}\right] \nonumber \\
&&  + {1\over {16\pi G_{5}}}\int_{\partial M}d^4 x \sqrt{-\gamma} 2K
\eear
Where K is the trace of the second fundamental form (external curvature) and $\gamma$ is the induced metric on the boundary. The Gibbons-Hawking boundary term in the second line is necessary for the consistency of Dirichlet problem for manifolds with boundary. In the coordinate system (\ref{holren1}) the equations of motion read as, \footnote{The convention for Riemann tensor that we follow is $R^{M}_{NPQ}=\partial_{P}\Gamma^{M}_{NQ}+\Gamma^{M}_{LP}\Gamma^{L}_{NQ}-P\leftrightarrow Q$ which is opposite to that of \cite{de Haro:2000xn}. Therefore there is a change in sign at various places in the equations of motion and in the analysis thereafter.}
\bear \label{holren3}
-{1\over 2}Tr(g^{-1}g^{\prime\prime})+{1\over 4}Tr(g^{-1}g^{\prime}g^{-1}g^{\prime})+{1\over 12}\left(-4\rho Tr(g^{-1}A^{\prime T}A^{\prime})-{1\over 4}Tr(g^{-1}Fg^{-1}F)\right)&=&0, \nonumber \\
 \nabla^{\mu}g^{\prime}_{\mu \nu}-\nabla_{\nu}Tr(g^{-1}g^{\prime})-\rho g^{\mu \sigma}A^{\prime}_{\mu}F_{\nu\sigma}&=&0, \nonumber \\
Ric[g]+\rho\left(-2g^{\prime\prime}+2 g^{\prime}g^{-1}g^{\prime}-Tr (g^{-1}g^{\prime})g^{\prime}\right)+2g^{\prime}+Tr(g^{-1}g^{\prime})g && \nonumber \\
+{1\over 12}g\left(8\rho^{2}A^{\prime T}g^{-1}A^{\prime}-\rho Tr(g^{-1}Fg^{-1}F)\right)+{1\over 2}\rho F g^{-1} F -2 \rho^{2}A^{\prime T}A^{\prime} &=& 0, \nonumber \\
{1 \over {\sqrt{-g}}} \partial_{\mu}\left(\sqrt{-g}g^{\mu\nu}A^{\prime}_{\nu}\right)-{{3\kappa}\over {8\sqrt{-g}}}\epsilon^{\mu\nu\sigma\delta}F_{\mu\nu}F_{\sigma\delta}&=&0, \nonumber \\
{{4 \rho}\over {\sqrt{-g}}}\partial_{\rho}\left(\sqrt{-g}g^{\mu\nu}A^{\prime}_{\nu}\right)-{1 \over {\sqrt{-g}}}\partial_{\sigma}\left(\sqrt{-g}(g^{-1}Fg^{-1})^{\mu\sigma}\right)-{{6\kappa \rho}\over {\sqrt{-g}}}\epsilon^{\mu\nu\sigma\delta}A^{\prime}_{\nu}F_{\sigma\delta}&=& 0. \nonumber \\
\eear
Differentiation with respect to $\rho$ is denoted by prime, $\nabla_{\mu}$ is the covariant derivative constructed from the metric $g_{\mu\nu}(x,\rho)$ and $Ric[g]$ is the Ricci tensor of the metric $g_{\mu\nu}$.\footnote{From now on we will suppress explicit mention of $g^{-1}$ in the terms. For example we will write $F^{2}_{\mu\nu}=(Fg^{-1}F)_{\mu\nu}$ and $Tr(F^2)=Tr(g^{-1}F g^{-1})F$ and so on.} We will now invoke the near boundary expansions as in (\ref{holren1}) in the above equations (\ref{holren3}). The solution we obtain are
\bear \label{holren4}
g^{(2)}_{\mu\nu}&=& -{1 \over 2}\left(R_{\mu\nu}-{R\over 6}g^{(0)}_{\mu\nu}\right), \quad \quad R_{\mu\nu}=Ric[g^{(0)}]\quad, \nonumber \\
h^{(4)}_{\mu\nu}&=& {1\over 8}\left(\nabla^{\sigma}\nabla_{\mu}g^{(2)}_{\sigma\nu}+\nabla^{\sigma}\nabla_{\nu}g^{(2)}_{\sigma\mu}-\nabla^{2}g^{(2)}_{\mu\nu}-\nabla_{\mu}\nabla_{\nu}Tr(g^{(2)})\right)+{1\over 2}g^{2}_{(2)\mu\nu}\nonumber \\
&&-{1\over 8}Tr(g^{2}_{(2)})g^{(0)}_{\mu\nu} -{1\over 32}Tr(F_{(0)}^{2})g^{(0)}_{\mu\nu}+ {1\over 8}F^{(0)2}_{\mu\nu} \nonumber \\
&=& {1\over 8}R_{\mu\sigma\nu\delta}R^{\sigma\delta}-{1\over 48}\nabla_{\mu}\nabla_{\nu}R+{1\over 16}\nabla^{2}R_{\mu\nu}-{1\over 24}R R_{\mu\nu}\nonumber \\
&& +\left(-{1\over 96}\nabla^{2}R + {1\over 96} R^2-{1\over 32}R_{\sigma\delta}R^{\sigma\delta}\right)g^{(0)}_{\mu\nu}-{1\over 32}Tr(F_{(0)}^{2})g^{(0)}_{\mu\nu}+ {1\over 8}F^{(0)2}_{\mu\nu} ,\nonumber \\
g^{(0)\mu\nu}B^{(2)}_{\nu}&=& {1\over 4}\nabla_{\nu}F^{(0)\mu\nu}, \nonumber \\
Tr(g_{(4)})&=& {1\over 4}Tr(g_{(2)}^{2})-{1\over 48}Tr(F_{(0)}^{2}), \nonumber \\
\nabla_{\mu}\left(g_{(0)}^{\mu\nu}\left(A_{\nu}^{(2)}+B_{\nu}^{(2)}\right)\right)&=& {{3\kappa}\over {8 \sqrt{-g^{(0)}}}}\epsilon^{\mu\nu\sigma\delta}F^{(0)}_{\mu\nu}F^{(0)}_{\sigma\delta}, \nonumber \\
\nabla^{\nu}g^{(4)}_{\mu\nu}&=& \nabla^{\nu}\left({1\over2}g_{(2)\mu\nu}^{2}-{1\over 4} (Tr g_{(2)})g^{(2)}_{\mu\nu}+{1\over 8}\left((Tr g_{(2)})^{2}-Tr(g_{(2)}^{2})\right)g^{(0)}_{\mu\nu}\right) \nonumber \\
&& -\nabla^{\nu}\left({1\over 48}(Tr(F_{(0)}^{2}))g^{(0)}_{\mu\nu}\right) +{1\over 2}g^{(0)\nu\sigma}\left(A^{(2)}_{\nu}+B^{(2)}_{\nu}\right)F^{(0)}_{\mu\sigma}. \nonumber \\
\eear
The last line in the above equation can be solved as
\be \label{holren5}
g^{(4)}_{\mu\nu}={1\over2}g_{(2)\mu\nu}^{2}-{1\over 4} (Tr g_{(2)})g^{(2)}_{\mu\nu}+{1\over 8}\left((Tr g_{(2)})^{2}-Tr(g_{(2)}^{2})\right)g^{(0)}_{\mu\nu}-{1\over 48}(Tr(F_{(0)}^{2}))g^{(0)}_{\mu\nu}+t_{\mu\nu}\,,
\ee
where $t_{\mu\nu}$ is arbitrary, but according to (\ref{holren4}), satisfies the following constraint
\bear \label{holren6}
\nabla^{\nu}t_{\mu\nu}&=& {1\over 2}g^{(0)\nu\sigma}\left(A^{(2)}_{\nu}+B^{(2)}_{\nu}\right)F^{(0)}_{\mu\sigma}, \nonumber \\
Tr t&=& {1\over 4}\left(Tr(g_{(2)}^{2})-(Tr g_{(2)})^2 + {1\over 4}(Tr F_{(0)}^{2})\right)\quad.
\eear

\subsection{Divergences and counter terms in terms of induced metric}

The on-shell gravitational action diverges near the boundary and hence we have to regulate the theory and remove potential divergence terms by adding local counter terms in the boundary. We follow the steps outlined in \cite{de Haro:2000xn}. We first regulate the theory by restricting the bulk integral to the region $\rho\geq \epsilon$. The regulated action is given by

\bear \label{holren7}
S_{reg} &=&{1\over {16\pi G_{5}}}\left[\int_{M} d^4 x d\rho \sqrt{-G}\left(R[G]+12- {1\over 4} F_{MN}F^{MN}\right) - {\kappa \over 4} \epsilon^{MNPQR}A_{M}F_{NP}F_{QR}\right] \nonumber \\
&&  + {1\over {16\pi G_{5}}}\int_{\partial M}d^4 x \sqrt{-\gamma} 2K, \nonumber \\
&=& {1\over {16\pi G_{5}}}\int d^4 x \left[-4\int_{\epsilon} {\sqrt{-g} \over \rho^{3}} d\rho -{1\over 12}\int_{\epsilon} {\sqrt{-g} \over \rho^{3}}F_{MN}F^{MN} d\rho - {\kappa \over 4} \int_{\epsilon} \epsilon^{MNPQR}A_{M}F_{NP}F_{QR} d\rho \right] \nonumber \\
&& +{1\over {16\pi G_{5}}}\int d^4 x \left[-{1\over\rho{2}}\left. \left(-8\sqrt{-g}+4\rho\partial_{\rho}\sqrt{-g}\right)\right|_{\rho=\epsilon}\right], \nonumber \\
&=& {1\over {16\pi G_{5}}}\int d^4 x \sqrt{-g_{(0)}}\left(6\epsilon^{-2}+a_{(4)}\log\epsilon\right)+\OO(\epsilon^{0})
\eear
Where $a_{(4)}={1\over2}\left[(Tr g_{(2)})^2-Tr(g_{(2)}^{2})-{1\over 4}Tr(F_{(0)}^{2})\right]$. In order to get a finite action we have to subtract the divergences by adding local counterterms in the boundary. For that we have to invert the relations between the induced metric $\gamma_{\mu\nu}={1\over \epsilon}g_{\mu\nu}$ and $g^{(0)}_{\mu\nu}$ perturbatively in $\epsilon$. The results
are \cite{de Haro:2000xn}
\bear \label{holren8}
\sqrt{-g_{(0)}}&=&\epsilon^{2}\left(1-{1\over 2}\epsilon Tr(g_{(2)})+\OO(\epsilon^{(2)})\right) \,, \nonumber \\
Tr(g_{(2)})&=&-{1\over {6\epsilon}}\left(R[\gamma])+\OO(R[\gamma]^{2})\right)\,, \nonumber \\
Tr(g_{(2)}^{2})&=& {1\over 4 \epsilon^2}\left(R_{\mu\nu}[\gamma]R^{\mu\nu}[\gamma]-{2\over 9}R^{2}[\gamma]+\OO(R[\gamma]^{3})\right)\,.
\eear
Using the above results the counter term action takes the following form
\be \label{holren9}
S^{ct}=-{1\over {16\pi G_5}}\int d^4 x ~ \sqrt{-\gamma}\left[6+{R[\gamma]\over 2}+a_4 [\gamma]\log \epsilon\right]
\ee
Where $a_4 [\gamma]$ is the same expression as given below (\ref{holren7}) with $g_{(0)}$ replaced by $\gamma$. While writing down the counter terms, we did not write down terms involving $R[\gamma]^2$ or $(\log\epsilon) R[\gamma]^{3}$ since these would yield finite contribution which would just be a different choice of scheme and will not affect physical quantitities like trace anomaly.

\subsection{Holographic stress energy tensor, current, and Ward identities} \label{}

Having found the counter terms required to cancel the divergences of the action in the previous section, we are in a position to derive the expressions for the expectation value of the stress energy tensor and current in the dual theory which are given as
\bear \label{holren10}
\langle T_{\mu\nu}[g_{(0)}] \rangle &=& {-2\over \sqrt{-g_{(0)}}}{{\delta S_{ren}}\over{\delta {g_{(0)}^{\mu\nu}}}}=\lim_{\epsilon \to 0}\left({1\over \epsilon}T_{\mu\nu}[\gamma]\right)\,, \nonumber \\
\langle J^{\mu}[g_{(0)}]\rangle &=& {1\over \sqrt{-g_{(0)}}}{{\delta S_{ren}}\over{\delta {A^{(0)}_{\mu}}}}=\lim_{\epsilon \to 0}\left({1\over {\epsilon^{2}}}J^{\mu}[\gamma]\right)\,,
\eear
where $T_{\mu\nu}[\gamma]$ and $J^{\mu}[\gamma]$ are the expressions for the stress energy tensor and current of the theory described at $\rho=\epsilon$ without taking the limit $\epsilon \to 0$. From the above expressions it is obvious that the expansions of $T_{\mu\nu}[\gamma]$ up to $\OO(\epsilon)$ and $J^{\mu}[\epsilon]$ up to $\OO(\epsilon^2)$ will be important. Both of these quantities have contributions from $S_{reg}$ as well as $S^{ct}$. The expressions coming from $S_{reg}$ are
\bear \label{holren11}
T_{\mu\nu}^{reg}[\gamma]&=&{1\over {8\pi G_5}}(K_{\mu\nu}-K\gamma_{\mu\nu})\nonumber\\&=&-{1\over {8\pi G_5}}\left(-\partial_{\epsilon} g_{\mu\nu}(x,\epsilon)+g_{\mu\nu}(x,\epsilon)Tr[g^{-1}(x,\epsilon)\partial_{\epsilon}g(x,\epsilon)]-{3\over \epsilon}g_{\mu\nu}(x,\epsilon)\right),  \\
J^{\mu}_{reg}&=& {\epsilon^{2}\over {8\pi G_5}}\left(\left(A^{(2)}_{\nu}+B^{(2)}_{\nu}\right)g_{(0)}^{\mu\nu}+g_{(0)}^{\mu\nu}B_{\nu}^{(2)}\log\epsilon-{\kappa \over {2\sqrt{-g_{(0)}}}}\epsilon^{\mu\nu\sigma\delta}A^{(0)}_{\nu}F^{(0)}_{\sigma\delta}\right)+\OO(\epsilon^3,\epsilon^3 \log\epsilon),\nonumber
\eear
The contribution coming from the counterterms (\ref{holren9}) are
\bear \label{holren12}
T_{\mu\nu}^{ct}&=& -{1\over {8\pi G_5}}\left(3\gamma_{\mu\nu}-{1\over 2}\left(R_{\mu\nu}[\gamma]-{1\over2}R[\gamma]\gamma_{\mu\nu}\right)-T^{a}_{\mu\nu}\log\epsilon\right), \nonumber \\
J^{\mu}_{ct}&=& {1\over {32 \pi G_5}}\nabla_{\nu}F^{\mu\nu}[\gamma]\log\epsilon = {\epsilon^2 \over 32 \pi G_5}\nabla_{\nu}F^{\mu\nu}[g_{(0)}]\log\epsilon\,,
\eear
where
\be \label{holren13}
T^{a}_{\mu\nu}={1\over\sqrt{-\gamma}}{{\delta\left(\int d^4 x \sqrt{-\gamma} a_{4}[\gamma]\right)}\over {\delta \gamma^{\mu\nu}}}=-2h^{(4)}_{\mu\nu}[\gamma]\quad.
\ee
In order to apply the above results to (\ref{holren10}), we need to expand them in powers of $\epsilon$, for which the following result will be useful,\footnote{The results are the same that appear in \cite{de Haro:2000xn} with slight change in signs of some terms because of the opposite convention used by us.}
\be \label{holren14}
R_{\mu\nu}[\gamma]=R_{\mu\nu}[g_{(0)}]+{1\over4}\epsilon \left( 2R_{\mu\sigma\nu\delta}R^{\sigma\delta}-2R_{\mu\sigma}R^{\sigma}_{\nu}-{1\over3}\nabla_{\mu}\nabla_{\nu}R+\nabla^{2}R_{\mu\nu}-{1\over6}(\nabla^2 R)g^{(0)}_{\mu\nu}\right).
\ee
After using this expression and after a slight bit of algebra, we observe that the $1\over\epsilon$ pole and logarithmic divergence in $\langle T_{\mu\nu}\rangle$ and the logarithmic divergence in $\langle J^{\mu} \rangle$ cancel and we get perfectly finite expressions for both of them which are as follows
\bear \label{holren15}
\langle T_{\mu\nu}\rangle &=& {1\over 8 \pi G_{5}}\left[2t_{\mu\nu}+3h^{(4)}_{\mu\nu}+{1\over4}\left({1\over4}(Tr F_{(0)}^2)g^{(0)}_{\mu\nu}-F^{(0)2}_{\mu\nu}\right)\right]\quad, \nonumber \\
\langle J^{\mu} \rangle &=& {1\over 8 \pi G_{5}}\left[g^{(0)\mu\nu}\left(A^{(2)}_{\nu}+B^{(2)}_{\nu}\right)-{\kappa\over2}\epsilon^{\mu\nu\sigma\delta}A^{(0)}_{\nu}F^{(0)}_{\sigma\delta}\right]
\quad.\eear
The expression for $\langle T_{\mu\nu}\rangle$ obtained above is almost the same as obtained in \cite{de Haro:2000xn} with the exception of the last two terms and explicit appearance of the boundary gauge field in the expression of $t_{\mu\nu}$ and $h^{(4)}_{\mu\nu}$. The trace anomaly that we obtain from the above expression is
\be \label{holren16}
Tr\langle T\rangle=-{1\over 8\pi G_5}a_{(4)}=-{1\over 16 \pi G_5}\left[(Tr g_{(2)})^2-Tr(g_{(2)}^{2})-{1\over 4}Tr(F_{(0)}^{2})\right]\quad.
\ee
Note that the terms proportional to $h^{(4)}_{\mu\nu}$ and ${1\over4}(Tr F_{(0)}^2)g^{(0)}_{\mu\nu}-F^{(0)2}_{\mu\nu}$ in $\langle T_{\mu\nu}\rangle$ are scheme dependent and can be removed by addition of local finite counterterms such as
\be
\int d^4 x \sqrt{-\gamma}\left[\alpha a_{(4)}[\gamma]+\beta Tr(F_{(0)}^{2})\right]\quad,
\ee
by appropriate choice of $\alpha$ and $\beta$. Indeed as expected they do not contribute to the trace anomaly. The trace anomaly now contains a term depending on the boundary gauge field and even in the flat space limit $g^{(0)}\to \eta$, we will still have a nonzero trace anomaly as expected. Removing the scheme dependent terms from the expectation value of the stress energy tensor and using the solutions to the equations of motion (\ref{holren4}) and (\ref{holren6}), we get the following Ward identities
\bear \label{holren17}
\nabla_{\nu}\langle T^{\mu\nu}\rangle &=& F^{(0)\mu\nu}\langle J_{\nu} \rangle -{\kappa\over 16\pi G_5 \sqrt{-g^{(0)}}}\epsilon^{\nu\sigma\alpha\beta}F^{(0)\mu}_{~\nu}A^{(0)}_{\sigma}F^{(0)}_{\alpha\beta} \,, \nonumber \\
\nabla_{\mu}\langle J^{\mu}\rangle &=& {\kappa\over 64 \pi G_5 \sqrt{-g^{(0)}}}\epsilon^{\mu\nu\sigma\delta}F^{(0)}_{\mu\nu}F^{(0)}_{\sigma\delta}\,.
\eear
Note the interesting contributions from the Chern-Simons term.

\section{Energy-momentum tensor and current of our solution}\label{sectionhol2}

After developing the necessary tools in the previous section, required to derive the energy momentum tensor and $U(1)$ current in the boundary from a bulk solution, we can apply it to our new gravity solutions up to next-leading order to find energy-momentum tensor and the current of our interest.
Some results in the previous section are covariant, especially the counter-terms (\ref{holren9}), so that they can be used in our Eddington-Finkelstein like coordinate in sections 2 and 3 as well.
In our situation, the boundary CFT metric $g^{(0)}$ is flat, so various pieces such as $g^{(2)}$ are simply absent, which makes things a lot simpler than it looks.
For completeness, we will state each steps of holographic renormalization in our coordinate system
explicitly, while main necessary computations can be simply borrowed from the previous section.

One first regularizes the action by considering only the bulk region of $r\le \epsilon^{-{1\over 2}}$ (note that $\rho$ in the previous section corresponds to $\rho=r^{-2}$ in our coordinate).
One can easily check that $\rho=\epsilon$ boundary in Fefferman-Graham coordinate is asymptotically identical to $r=\epsilon^{-{1\over 2}}$ boundary in our Eddington-Finkelstein coordinate, so that the covariant counter-terms in the previous section can be used without any modification.
The explicit expression for the counter-terms is given by (\ref{holren9}),
\be
S^{ct}=-{1\over 16\pi G_5}\int d^4 x\, \sqrt{-\gamma}\left(6+{1\over 2}R[\gamma]+a_4[\gamma]\log\epsilon\right)\quad,
\ee
from which the counter-term contribution to the energy-momentum tensor is
\be
T^{ct}_{\mu\nu}=-{1\over 8\pi G_5} \left(3\gamma_{\mu\nu}-{1\over 2}\left(R_{\mu\nu}(\gamma)-{1\over 2} R(\gamma) \gamma_{\mu\nu}\right) +2 h^{(4)}_{\mu\nu} \log\epsilon \right)\quad,
\ee
with $h^{(4)}$ now becomes, using the flat CFT metric $g^{(0)}=\eta$,
\be
h^{(4)}_{\mu\nu}={1\over 32}\left(F^{(0)}_{\alpha\beta}F^{(0)\alpha\beta}\right)\eta_{\mu\nu} +{1\over 8} F^{(0)}_{\mu\alpha}F^{(0)\alpha}_{\quad\,\,\,\,\, \nu}\quad,\label{h4}
\ee
where $\gamma$ is the induced metric on the boundary, and $A_\mu^{(0)}$ is the external gauge potential coupled to the U(1) current. In (\ref{h4}), indices are raised/lowered by $g^{(0)}=\eta$.
The regularized bulk contribution to the energy-momentum tensor is given by the Brown-York expression even in the presence of Maxwell field, \footnote{This is because using the bulk equations of motion, the variation of bulk action with respect to metric variation always reduces to the same surface boundary term which is the Brown-York tensor. Maxwell-CS term does not produce any boundary surface term with respect to metric variation.}
\be
T^{reg}_{\mu\nu}={1\over 8\pi G_5} \left(K_{\mu\nu}-K \gamma_{\mu\nu}\right)\quad,
\ee
which is already covariant so that we can also use this in our coordinate system.
It is then straightforward to compute finite, renormalized energy-momentum tensor in our coordinate system using the above expressions,
\be
T^{ren}_{\mu\nu}=\lim_{\epsilon\to 0}{1\over\epsilon}\left(T^{reg}_{\mu\nu}(\epsilon)+T^{ct}_{\mu\nu}(\epsilon)\right)\quad.
\ee

After plugging in the general, spatially-homogeneous near-boundary solutions of (\ref{nbdry}) in the presence of external electric field that we reproduce below
\bear
a(r,t)&=& 1+{a^{(1)}(t)\over r}+\left({1\over 4}\left(a^{(1)}(t)\right)^2 -\partial_t a^{(1)}(t)\right) {1\over r^2}
+{a^{(4)}(t)\over r^4}-{E^2\over 6}{\log r\over r^4}+\cdots\,,\nonumber\\
b(r,t)&=& 1+{a^{(1)}(t)\over r}+{\left(a^{(1)}(t)\right)^2\over 4} {1\over r^2}
+{b^{(4)}(t)\over r^4}-{E^2\over 6}{\log r\over r^4}+\cdots\,,\nonumber\\
c(r,t)&=& 1+{a^{(1)}(t)\over r}+{\left(a^{(1)}(t)\right)^2\over 4} {1\over r^2}
+\left(-{1\over 2}b^{(4)}(t)-{E^2\over 24}\right){1\over r^4}+{E^2\over 12}{\log r\over r^4}+\cdots\,,\nonumber\\
A(r,t)&=& E t + {E\over r}+ \left(j^{(2)}(t)-{1\over 2}E a^{(1)}(t)\right){1\over r^2}  + 0\cdot {\log r\over r^2} +\cdots\,,\label{nbdry2}
\eear
one obtains indeed the finite, well-defined CFT energy-momentum tensor as
\bear
T^{tt}&=&{1\over 8\pi G_5}\left(-{3\over 2}a^{(4)}(t)-{1\over 6} E^2\right)\quad,\nonumber\\
T^{zz}&=& {1\over 8\pi G_5}\left(-{1\over 2} a^{(4)}(t)+2 b^{(4)}(t)+{1\over 6}E^2\right)\quad,\nonumber\\
T^{xx}=T^{yy}&=&{1\over 8\pi G_5}\left(-{1\over 2} a^{(4)}(t) -b^{(4)}(t)-{1\over 24} E^2\right)\quad,
\eear
with dynamically unknown functions $a^{(4)}$ and $b^{(4)}$ that can be read off from our explicit solutions in section 3. As claimed before, $a^{(1)}(t)$ never appears in the results because it is a spurious mode of coordinate re-parametrization. Note that the above result nicely confirms the conformal anomaly (\ref{holren16}) we
obtain in the previous section,
\be
T^{\mu}_{\,\,\,\mu} = {1\over 32\pi G_5} E^2 = -{1\over 64\pi G_5} \label{traceano} F^{(0)}_{\mu\nu}F^{(0)\mu\nu}\quad,
\ee
which is a consistency check. Due to this constraint, one can't completely ignore contact terms of $E^2$ in the energy-momentum tensor above. However, as long as it satisfies the trace anomaly (\ref{traceano}), one is allowed to move them from place to place, which are different scheme choices.
A similar procedure can be followed to obtain the renormalized value of U(1) current; note that the same counter term (\ref{holren9}) also takes care of divergences from the Maxwell(-CS) action.
The result is simply
\be
J_z={1\over 8\pi G_5} j^{(2)}(t)\quad,
\ee
where $j^{(2)}$ is another dynamical function appearing in (\ref{nbdry2}). From the constraint (\ref{curcons}) dictated by the equations of motion
\be
\partial_t a^{(4)}(t)={2\over 3} E \,j^{(2)}(t)\quad,\label{curcons2}
\ee
one also checks the Ward identity (\ref{holren17}) to hold,
\be
\partial_{\nu}T^{\mu\nu}= F^{(0)\mu\nu} J_{\nu} \quad,
\ee
where the Chern-Simons piece in (\ref{holren17}) is not relevant in our specific situation.

What remains to obtain final results of $(T_{\mu\nu},J_\mu)$ is to simply retrieve the necessary dynamical functions $(a^{(4)},b^{(4)},j^{(2)})$ from our explicit solutions in section 3, and
it is not difficult to get
\bear
a^{(4)}(t)&=& -\left(E\over 2\right)^{8\over 3} t^{4\over 3}+{E^2\over 18}\left(\log t+2\log E+6\left(\log 2\right)^2-{1\over 2}\log 2 -2\right)+{\cal O}\left(t^{-{4\over 3}}\right)\quad,\nonumber\\
b^{(4)}(t)&=& {E^2\over 18}\left(\log t +2 \log E -{1\over 2}\log 2 +{7\over 4}\right)+{\cal O}\left(t^{-{4\over 3}}\right)\quad,\nonumber\\
j^{(2)}(t)&=&-\left(E\over 2\right)^{5\over 3} t^{1\over 3} +{E\over 12} \,t^{-1} +{\cal O}\left(t^{-{7\over 3}}\right)\quad.
\eear
These are the main results in this paper. In particular we see that the $T^{tt}$ component of the energy momentum tensor increases with time \footnote{Note that the solution presented is reliable only in the late time regimes as we have worked out the solution in a late time expansion scheme (\ref{newexp1},\ref{newexp2})} which is consistent with our expectation of energy being pumped into the system by an external electric field and the rate of increase of energy density being pumped into the system is also consistent with the Ward identity arising out of a rigorous holographic renormalization presented in section \ref{sectionhol} taking into account the full backreaction of the gravity solution. We also see that the energy density $T^{tt}$ has the correct $T^4$ ($T$ being the temperature) or $t^{4/3}$ behaviour in the leading order as expected from a conformal plasma but what is non trivial are the subleading terms which are the new results of our analysis.

\section{Discussion and future direction}

Time dependent gravity background is a valuable play ground to check several interesting dynamical phenomena in gravity. It becomes even more interesting for the asymptotic AdS gravity as it
holographically describes a dual conformal field theory at strong coupling whose study is typically
intractable by conventional field theory techniques. Lattice studies for time-dependent dynamical aspects also possess intrinsic problems. In this paper, we provide one more example of "quasi-analytic" time-dependent AdS gravity solution by applying external electric field to the system of U(1) symmetry, which pumps in energy flow into the system so that the plasma can be heated up continuously.
It is "quasi-analytic" because the solution is obtained by invoking a late-time expansion scheme we propose, from which one can construct solutions order by order. We check consistency of the scheme and provide analytic solutions up to next-leading order, from which we obtain late-time behaviors of energy-momentum tensor and U(1) current. For this purpose we also perform rigorous holographic renormalization of 5D asymptotic AdS Einstein-Maxwell(-Chern-Simons) theory considering full back-reactions to the metric. Note that back-reactions should be included to find precise modifications of energy-momentum Ward identities due to U(1) symmetry. We find our results for the current density and energy momentum tensor to be consistent with the Ward identity derived in section \ref{sectionhol}. We also obtain the correct leading behaviour $\sim T^4$ for the energy density and also obtain non trivial subleading behaviours which are the new results in this paper.  We would also like to stress that the results of section 4, on the modification of conformal Ward identities and trace anomaly in the presence of a background gauge field, are new and interesting on its own irrespective of the other results of the paper. Specifically we performed a rigorous holographic renormalization procedure on the Einstein-Maxwell-CS system by taking into account full backreactions from the metric and the results (\ref{holren16}, \ref{holren17}) gives us the complete modified conformal Ward identity and trace anomaly in the presence of a non trivial background gauge field. The appearance of the coefficient of the Chern Simons term $\kappa$ in the above results seems to encompass the effects from triangle anomaly in the field theory side and it would be interesting to study this effect by taking into account background magnetic field. We will discuss on this open issue more towards the end.

We close this section by a few questions regarding future directions.

{\it Apparent horizon and entropy density :}
It is not the area of event horizon, but that of  apparent horizon which gives one the entropy
density in the dual gravity plasma \cite{Bhattacharyya:2008xc}. This distinction is important in a time-dependent background.
It will be interesting to compute entropy density as a function of time by identifying apparent horizon of our new geometry.

{\it Hydrodynamic interpretation  with external electric field :}
Because the ratio $E\over T^2$ becomes arbitrary small in the late time regime, one expects that our dual plasma may be described by some form of hydrodynamics with electric field. Especially interesting question is whether our next-leading solution gives some kinds of transport coefficients. However, hydrodynamics is normally based on spatial-gradient expansion, while our plasma simply doesn't have any spatial-gradients because we are assuming translation invariance. This might imply that no hydrodynamic modes are excited at all, and
non-hydrodynamic modes might be governing the dynamics of our plasma. The situation seems similar to Ref.\cite{Chesler:2008hg}\footnote{We thank Andy O'Bannon for a discussion on this.}.

{\it Boost-invariant symmetry and early-time thermalization study :}
Instead of assuming translation symmetry that we study in this paper, one can also impose boost-invariance on the system with external electric field. It simply corresponds to a different preparation/initial condition of the system. Boost-invariance frame naturally causes a cooling tendency due to metric expansion while electric field will try to heat up the plasma, so there can be interesting interplay between the two effects.
Another advantage of having boost-invariance rather than translation-invariance is that a boost-invariant plasma is guaranteed to have hydrodynamic modes excited, so the question of hydrodynamic description can be more transparent \cite{Chesler:2009cy}.
Independently, it is also interesting to study early-time thermalization process starting at zero temperature and applying electric field after $t=0$. These will be addressed in a near future \cite{andy}.

{\it Charged plasma with electric field :}
Although we restrict our focus on a neutral plasma in this work, it will be an interesting extension to consider a charged plasma with external electric field. The additional complication will be that the electric field causes the net momentum flow along its direction, and one can not remove $g_{ti}$-components in the 5D metric ansatz in any frame. Even 4D boost-transformations cannot completely remove them from the bulk 5D metric because charge conjugation symmetry no longer prohibits them.

{\it Both electric and magnetic fields, and effects from triangle anomaly :}
The case with both electric and magnetic field pointing to a same direction seems particularly interesting in regard to effects from triangle anomaly \cite{Domokos:2007kt,Erdmenger:2008rm,Lifschytz:2009si,Son:2009tf,Fukushima:2008xe,Yee:2009vw,Matsuo:2009xn,Chuang:2010ku}; there will be a constant anomalous creation of charge density due to electro-magnetic fields, in addition to current/momentum flows.
As the 5-dimensional Chern-Simons term is holographic dual to the triangle anomaly, it will play an essential role in this case. On a practical side, what one needs is a more general ansatz in the gravity side including $(A_t,g_{ti})$ as well, while $F_{12}\equiv B$ will be a constant in full 5D due to Bianchi identity. It seems to be a tractable problem to our eyes.

\vskip 1cm \centerline{\large \bf Acknowledgement} \vskip 0.5cm

We greatly appreciate early collaboration and many useful discussions with Andy O'Bannon.
This project was initiated by his comment in a talk at SISSA, 2009 winter, on the necessity of having expanding horizon in the presence of electric field. We also thank Amos Yarom for helpful discussion

%%%%%%%%%%%%%%%%%%%%%%%%%%%%%%%%%%%%%%%%%%%%%%%%%%%%%%%%%%%%%%%%%%%
 \vfil


\begin{thebibliography}{99} \frenchspacing



%\cite{Maldacena:1997re}
\bibitem{Maldacena:1997re}
  J.~M.~Maldacena,
  ``The large N limit of superconformal field theories and supergravity,''
  Adv.\ Theor.\ Math.\ Phys.\  {\bf 2}, 231 (1998)
  [Int.\ J.\ Theor.\ Phys.\  {\bf 38}, 1113 (1999)]
  [arXiv:hep-th/9711200].
  %%CITATION = IJTPB,38,1113;%%

%\cite{Witten:1998zw}
\bibitem{Witten:1998zw}
  E.~Witten,
  ``Anti-de Sitter space, thermal phase transition, and confinement in  gauge
  theories,''
  Adv.\ Theor.\ Math.\ Phys.\  {\bf 2}, 505 (1998)
  [arXiv:hep-th/9803131].
  %%CITATION = 00203,2,505;%%

%\cite{Policastro:2002se}
\bibitem{Policastro:2002se}
  G.~Policastro, D.~T.~Son and A.~O.~Starinets,
  ``From AdS/CFT correspondence to hydrodynamics,''
  JHEP {\bf 0209}, 043 (2002)
  [arXiv:hep-th/0205052].
  %%CITATION = JHEPA,0209,043;%%


%\cite{Shuryak:2003ty}
\bibitem{Shuryak:2003ty}
  E.~V.~Shuryak and I.~Zahed,
  ``Rethinking the properties of the quark gluon plasma at T approx. T(c),''
  Phys.\ Rev.\  C {\bf 70}, 021901 (2004)
  [arXiv:hep-ph/0307267];\\
  %%CITATION = PHRVA,C70,021901;%%
%\cite{Teaney:2003kp}
%\bibitem{Teaney:2003kp}
  D.~Teaney,
  ``Effect of shear viscosity on spectra, elliptic flow, and Hanbury
  Brown-Twiss radii,''
  Phys.\ Rev.\  C {\bf 68}, 034913 (2003)
  [arXiv:nucl-th/0301099].
  %%CITATION = PHRVA,C68,034913;%%

%\cite{Kovtun:2003wp}
\bibitem{Kovtun:2003wp}
  P.~Kovtun, D.~T.~Son and A.~O.~Starinets,
  ``Holography and hydrodynamics: Diffusion on stretched horizons,''
  JHEP {\bf 0310}, 064 (2003)
  [arXiv:hep-th/0309213].
  %%CITATION = JHEPA,0310,064;%%

%\cite{Buchel:2003tz}
\bibitem{Buchel:2003tz}
  A.~Buchel and J.~T.~Liu,
  ``Universality of the shear viscosity in supergravity,''
  Phys.\ Rev.\ Lett.\  {\bf 93}, 090602 (2004)
  [arXiv:hep-th/0311175].
  %%CITATION = PRLTA,93,090602;%%


%\cite{Bhattacharyya:2008jc}
\bibitem{Bhattacharyya:2008jc}
  S.~Bhattacharyya, V.~E.~Hubeny, S.~Minwalla and M.~Rangamani,
  ``Nonlinear Fluid Dynamics from Gravity,''
  JHEP {\bf 0802}, 045 (2008)
  [arXiv:0712.2456 [hep-th]].
  %%CITATION = JHEPA,0802,045;%%

%\cite{Natsuume:2007ty}
\bibitem{Natsuume:2007ty}
  M.~Natsuume and T.~Okamura,
  ``Causal hydrodynamics of gauge theory plasmas from AdS/CFT duality,''
  Phys.\ Rev.\  D {\bf 77}, 066014 (2008)
  [Erratum-ibid.\  D {\bf 78}, 089902 (2008)]
  [arXiv:0712.2916 [hep-th]].
  %%CITATION = PHRVA,D77,066014;%%


%\cite{Iqbal:2008by}
\bibitem{Iqbal:2008by}
  N.~Iqbal and H.~Liu,
  ``Universality of the hydrodynamic limit in AdS/CFT and the membrane
  paradigm,''
  Phys.\ Rev.\  D {\bf 79}, 025023 (2009)
  [arXiv:0809.3808 [hep-th]].
  %%CITATION = PHRVA,D79,025023;%%

%\cite{Son:2007vk}
\bibitem{Son:2007vk}
  D.~T.~Son and A.~O.~Starinets,
  ``Viscosity, Black Holes, and Quantum Field Theory,''
  Ann.\ Rev.\ Nucl.\ Part.\ Sci.\  {\bf 57}, 95 (2007)
  [arXiv:0704.0240 [hep-th]].
  %%CITATION = ARNUA,57,95;%%

%\cite{Rangamani:2009xk}
\bibitem{Rangamani:2009xk}
  M.~Rangamani,
  ``Gravity \& Hydrodynamics: Lectures on the fluid-gravity correspondence,''
  Class.\ Quant.\ Grav.\  {\bf 26}, 224003 (2009)
  [arXiv:0905.4352 [hep-th]].
  %%CITATION = CQGRD,26,224003;%%

%\cite{Eling:2010vr}
\bibitem{Eling:2010vr}
  C.~Eling, I.~Fouxon and Y.~Oz,
  ``Gravity and a Geometrization of Turbulence,''
  arXiv:1004.2632 [Unknown].
  %%CITATION = ARXIV:1004.2632;%%


%\cite{Herzog:2002pc}
\bibitem{Herzog:2002pc}
  C.~P.~Herzog and D.~T.~Son,
  ``Schwinger-Keldysh propagators from AdS/CFT correspondence,''
  JHEP {\bf 0303}, 046 (2003)
  [arXiv:hep-th/0212072].
  %%CITATION = JHEPA,0303,046;%%

%\cite{Gubser:1998bc}
\bibitem{Gubser:1998bc}
  S.~S.~Gubser, I.~R.~Klebanov and A.~M.~Polyakov,
  ``Gauge theory correlators from non-critical string theory,''
  Phys.\ Lett.\  B {\bf 428}, 105 (1998)
  [arXiv:hep-th/9802109];\\
  %%CITATION = PHLTA,B428,105;%%
%\cite{Witten:1998qj}
%\bibitem{Witten:1998qj}
  E.~Witten,
  ``Anti-de Sitter space and holography,''
  Adv.\ Theor.\ Math.\ Phys.\  {\bf 2}, 253 (1998)
  [arXiv:hep-th/9802150].
  %%CITATION = 00203,2,253;%%


%\cite{de Haro:2000xn}
\bibitem{de Haro:2000xn}
  S.~de Haro, S.~N.~Solodukhin and K.~Skenderis,
  ``Holographic reconstruction of spacetime and renormalization in the  AdS/CFT
  correspondence,''
  Commun.\ Math.\ Phys.\  {\bf 217}, 595 (2001)
  [arXiv:hep-th/0002230].
  %%CITATION = CMPHA,217,595;%%




%\cite{Kovchegov:2007pq}
\bibitem{Kovchegov:2007pq}
  Y.~V.~Kovchegov and A.~Taliotis,
  ``Early time dynamics in heavy ion collisions from AdS/CFT correspondence,''
  Phys.\ Rev.\  C {\bf 76}, 014905 (2007)
  [arXiv:0705.1234 [hep-ph]].
  %%CITATION = PHRVA,C76,014905;%%

%\cite{Gubser:2008pc}
\bibitem{Gubser:2008pc}
  S.~S.~Gubser, S.~S.~Pufu and A.~Yarom,
  ``Entropy production in collisions of gravitational shock waves and of heavy
  ions,''
  Phys.\ Rev.\  D {\bf 78}, 066014 (2008)
  [arXiv:0805.1551 [hep-th]].
  %%CITATION = PHRVA,D78,066014;%%


%\cite{Chesler:2008hg}
\bibitem{Chesler:2008hg}
  P.~M.~Chesler and L.~G.~Yaffe,
  ``Horizon formation and far-from-equilibrium isotropization in supersymmetric
  Yang-Mills plasma,''
  Phys.\ Rev.\ Lett.\  {\bf 102}, 211601 (2009)
  [arXiv:0812.2053 [hep-th]].
  %%CITATION = PRLTA,102,211601;%%

%\cite{Lin:2009pn}
\bibitem{Lin:2009pn}
  S.~Lin and E.~Shuryak,
  ``Grazing Collisions of Gravitational Shock Waves and Entropy Production in
  Heavy Ion Collision,''
  Phys.\ Rev.\  D {\bf 79}, 124015 (2009)
  [arXiv:0902.1508 [hep-th]].
  %%CITATION = PHRVA,D79,124015;%%

%\cite{Bhattacharyya:2009uu}
\bibitem{Bhattacharyya:2009uu}
  S.~Bhattacharyya and S.~Minwalla,
  ``Weak Field Black Hole Formation in Asymptotically AdS Spacetimes,''
  JHEP {\bf 0909}, 034 (2009)
  [arXiv:0904.0464 [hep-th]].
  %%CITATION = JHEPA,0909,034;%%

%\cite{Chesler:2009cy}
\bibitem{Chesler:2009cy}
  P.~M.~Chesler and L.~G.~Yaffe,
  ``Boost invariant flow, black hole formation, and far-from-equilibrium
  dynamics in N = 4 supersymmetric Yang-Mills theory,''
  arXiv:0906.4426 [hep-th].
  %%CITATION = ARXIV:0906.4426;%%

\bibitem{irina}
I.~Ya.~Aref'eva, A.~A.~Bagrov and L.~V.~Joukovskaya,
``Critical Trapped Surfaces Formation in the Collision of
Ultrarelativistic Charges,''
arXiv:0909.1294.



%\cite{CaronHuot:2006te}
\bibitem{CaronHuot:2006te}
  S.~Caron-Huot, P.~Kovtun, G.~D.~Moore, A.~Starinets and L.~G.~Yaffe,
  ``Photon and dilepton production in supersymmetric Yang-Mills plasma,''
  JHEP {\bf 0612}, 015 (2006)
  [arXiv:hep-th/0607237].
  %%CITATION = JHEPA,0612,015;%%

%\cite{Karch:2007pd}
\bibitem{Karch:2007pd}
  A.~Karch and A.~O'Bannon,
  ``Metallic AdS/CFT,''
  JHEP {\bf 0709}, 024 (2007)
  [arXiv:0705.3870 [hep-th]].
  %%CITATION = JHEPA,0709,024;%%


  %\cite{Janik:2005zt}
\bibitem{Janik:2005zt}
  R.~A.~Janik and R.~B.~Peschanski,
  ``Asymptotic perfect fluid dynamics as a consequence of AdS/CFT,''
  Phys.\ Rev.\  D {\bf 73}, 045013 (2006)
  [arXiv:hep-th/0512162].
  %%CITATION = PHRVA,D73,045013;%%


%\cite{D'Hoker:2009mm}
\bibitem{D'Hoker:2009mm}
  E.~D'Hoker and P.~Kraus,
  ``Magnetic Brane Solutions in AdS,''
  JHEP {\bf 0910}, 088 (2009)
  [arXiv:0908.3875 [hep-th]];\\
  %%CITATION = JHEPA,0910,088;%%
%\cite{D'Hoker:2009bc}
%\bibitem{D'Hoker:2009bc}
  E.~D'Hoker and P.~Kraus,
  ``Charged Magnetic Brane Solutions in $AdS_5$ and the fate of the third law of
  thermodynamics,''
  JHEP {\bf 1003}, 095 (2010)
  [arXiv:0911.4518 [Unknown]].
  %%CITATION = JHEPA,1003,095;%%

%\cite{Kinoshita:2008dq}
\bibitem{Kinoshita:2008dq}
  S.~Kinoshita, S.~Mukohyama, S.~Nakamura and K.~y.~Oda,
  ``A Holographic Dual of Bjorken Flow,''
  Prog.\ Theor.\ Phys.\  {\bf 121}, 121 (2009)
  [arXiv:0807.3797 [hep-th]].
  %%CITATION = PTPKA,121,121;%%


%\cite{Bhattacharyya:2008xc}
\bibitem{Bhattacharyya:2008xc}
  S.~Bhattacharyya {\it et al.},
  ``Local Fluid Dynamical Entropy from Gravity,''
  JHEP {\bf 0806}, 055 (2008)
  [arXiv:0803.2526 [hep-th]].
  %%CITATION = JHEPA,0806,055;%%

\bibitem{andy}
A.~O'Bannon, B.~Sahoo, H.~-U.~Yee, work in progress.

%\cite{Domokos:2007kt}
\bibitem{Domokos:2007kt}
  S.~K.~Domokos and J.~A.~Harvey,
  ``Baryon number-induced Chern-Simons couplings of vector and axial-vector
  mesons in holographic QCD,''
  Phys.\ Rev.\ Lett.\  {\bf 99}, 141602 (2007)
  [arXiv:0704.1604 [hep-ph]].
  %%CITATION = PRLTA,99,141602;%%

\bibitem{Erdmenger:2008rm}
  J.~Erdmenger, M.~Haack, M.~Kaminski and A.~Yarom,
  ``Fluid dynamics of R-charged black holes,''
  JHEP {\bf 0901}, 055 (2009),
  [arXiv:0809.2488 [hep-th]];\\
  %%CITATION = JHEPA,0901,055;%%
%\cite{Banerjee:2008th}
%\bibitem{Banerjee:2008th}
  N.~Banerjee, J.~, S.~Bhattacharyya, S.~Dutta, R.~Loganayagam and P.~Surowka,
  ``Hydrodynamics from charged black branes,''
  arXiv:0809.2596 [hep-th];\\
  %\cite{Hur:2008tq}
%\bibitem{Hur:2008tq}
  J.~Hur, K.~K.~Kim and S.~J.~Sin,
  ``Hydrodynamics with conserved current from the gravity dual,''
  JHEP {\bf 0903}, 036 (2009)
  [arXiv:0809.4541 [hep-th]];\\
  %%CITATION = JHEPA,0903,036;%%
  %%CITATION = ARXIV:0809.2596;%%
%\cite{Torabian:2009qk}
%\bibitem{Torabian:2009qk}
  M.~Torabian and H.~U.~Yee,
  ``Holographic nonlinear hydrodynamics from AdS/CFT with multiple/non-Abelian
  symmetries,''
  JHEP {\bf 0908}, 020 (2009),
  [arXiv:0903.4894 [hep-th]].
  %%CITATION = JHEPA,0908,020;%%




%\cite{Lifschytz:2009si}
\bibitem{Lifschytz:2009si}
  G.~Lifschytz and M.~Lippert,
  ``Anomalous conductivity in holographic QCD,''
  Phys.\ Rev.\  D {\bf 80}, 066005 (2009)
  [arXiv:0904.4772 [hep-th]].
  %%CITATION = PHRVA,D80,066005;%%


%\cite{Son:2009tf}
\bibitem{Son:2009tf}
  D.~T.~Son and P.~Surowka,
  ``Hydrodynamics with Triangle Anomalies,''
  Phys.\ Rev.\ Lett.\  {\bf 103}, 191601 (2009)
  [arXiv:0906.5044 [hep-th]].
  %%CITATION = PRLTA,103,191601;%%

%\cite{Fukushima:2008xe}
\bibitem{Fukushima:2008xe}
  K.~Fukushima, D.~E.~Kharzeev and H.~J.~Warringa,
  ``The Chiral Magnetic Effect,''
  Phys.\ Rev.\  D {\bf 78}, 074033 (2008).
  [arXiv:0808.3382 [hep-ph]].
  %%CITATION = PHRVA,D78,074033;%%

%\cite{Yee:2009vw}
\bibitem{Yee:2009vw}
  H.~U.~Yee,
  ``Holographic Chiral Magnetic Conductivity,''
  JHEP {\bf 0911}, 085 (2009)
  [arXiv:0908.4189 [hep-th]];\\
  %%CITATION = JHEPA,0911,085;%%
%\cite{Rebhan:2009vc}
%\bibitem{Rebhan:2009vc}
  A.~Rebhan, A.~Schmitt and S.~A.~Stricker,
  ``Anomalies and the chiral magnetic effect in the Sakai-Sugimoto model,''
  JHEP {\bf 1001}, 026 (2010)
  [arXiv:0909.4782 [hep-th]];\\
  %%CITATION = JHEPA,1001,026;%%
%\cite{Gorsky:2010xu}
%\bibitem{Gorsky:2010xu}
  A.~Gorsky, P.~N.~Kopnin and A.~V.~Zayakin,
  ``On the Chiral Magnetic Effect in Soft-Wall AdS/QCD,''
  arXiv:1003.2293 [hep-ph].
  %%CITATION = ARXIV:1003.2293;%%



%\cite{Matsuo:2009xn}
\bibitem{Matsuo:2009xn}
  Y.~Matsuo, S.~J.~Sin, S.~Takeuchi and T.~Tsukioka,
  ``Magnetic conductivity and Chern-Simons Term in Holographic Hydrodynamics of
  Charged AdS Black Hole,''
  arXiv:0910.3722 [hep-th];\\
  %%CITATION = ARXIV:0910.3722;%%
%\cite{Sahoo:2009yq}
%\bibitem{Sahoo:2009yq}
  B.~Sahoo and H.~U.~Yee,
  ``Holographic chiral shear waves from anomaly,''
  arXiv:0910.5915 [hep-th];\\
  %%CITATION = ARXIV:0910.5915;%%
%\cite{Nakamura:2009tf}
%\bibitem{Nakamura:2009tf}
  S.~Nakamura, H.~Ooguri and C.~S.~Park,
  ``Gravity Dual of Spatially Modulated Phase,''
  Phys.\ Rev.\  D {\bf 81}, 044018 (2010)
  [arXiv:0911.0679 [hep-th]].
  %%CITATION = PHRVA,D81,044018;%%

%\cite{Chuang:2010ku}
\bibitem{Chuang:2010ku}
  W.~y.~Chuang, S.~H.~Dai, S.~Kawamoto, F.~L.~Lin and C.~P.~Yeh,
  ``Dynamical Instability of Holographic QCD at Finite Density,''
  arXiv:1004.0162 [Unknown].
  %%CITATION = ARXIV:1004.0162;%%


\end{thebibliography}
\end{document}